\documentclass[10pt]{article}
\usepackage{amssymb}
\usepackage[numbers,compress]{natbib}
\begin{document}

\title{Discrete symmetries of  Dirac's theory in the de Sitter manifold}

\author{Ion I. Cot\u aescu\thanks{Corresponding author E-mail:~~i.cotaescu@e-uvt.ro}~  and  Ion Cot\u aescu Jr.\thanks{E-mail:~~ion.cotaescu@e-uvt.ro}\\
{\it West University of Timi\c soara,} \\{\it V. Parvan Ave. 4,
RO-300223 Timi\c soara}}

\maketitle

\begin{abstract}
The discrete symmetries of the Dirac field on the de Sitter manifold are studied taking into account that this has two portions that can play the role of  physical space-times, namely the expanding and a collapsing universes. The proper discrete  isometries which preserve the portion have a physical meaning in contrast to the improper ones which change the portion remaining thus of a mere mathematical interest. The discrete symmetries generated by the proper isometries and charge conjugation are studied in physical frames on the expanding portion shoving that all the discrete transformations reversing the cosmic time are local depending on a local boost matrix. Thus  the  discrete group of Dirac's theory in the de Sitter expanding universe  is obtained showing that this is of the order 16 having a multiplication table similar to that of Dirac's theory in special relativity.  Moreover, all the discrete de Sitter isometries, including the improper ones, are studied in conformal frames for obtaining a global image about the de Sitter isometries in spite of the fact that these  cannot be gathered in a larger discrete group with physical meaning. 
\end{abstract}

Pacs: 04.62.+v

\vspace*{12mm} Keywords: de Sitter space-time; discrete symmetries; proper isometries; improper isometries; discrete symmetry group.  

\newpage
\section{Introduction}

The Dirac field on the $(3+1)$-dimensional de Sitter manifold, $M$, embedded in a $(1+4)$-dimensional pseudo-Euclidean manifold, $M^5$, was considered in two dissimilar manners, namely either as an {\em invariant} field defined on $M^5$ or as a {\em covariant} one defined on $M$. The first approach referred often as the de Sitter ambient was developed by many authors considering a spinor field transforming under isometries according to linear representations  of the universal covering group ${\rm Spin}SO(1,4)$ of the stable group $SO(1,4)$ of the embedding space $M^5$  (see for instance \cite{Gaz,G,T,V}). 

The theory of the covariant Dirac field defined on $M$ was initiated by Nachtmann \cite{Nach} and developed then by one of us \cite{ES,CD1,CSchr,CD2,Ccov,Cquant}. In this approach  the  Dirac field transforms under continuous   $O(1,4)$ isometries according to the Dirac representation $\rho_D$ of the  $SL(2,\Bbb{C})={\rm Spin}L_+^{\uparrow}$ group  which is the universal covering group of the proper orthochronus Lorentz group $L_+^{\uparrow}$ that is {\em gauge} group of $M$ \cite{WKT}.  The $SL(2,\Bbb{C})$ transformations are associated through the canonical homomorphism  to the gauge ones  which have to correct  the tetrad gauge after each isometry in order to leave the Dirac equation invariant \cite{ES} but giving rice to the associated conserved quantities predicted by Carter and McLenaghan \cite{CMLa,EPL,GRG}. For this reason we say that the covariant representation of the isometry group $SO(1,4)$ is {\em induced}  by $\rho_D$. In other words  this approach extends the Lorentz covariance of special relativity to general relativity. This framework offers many attractive applications \cite{A1,A2,A3,A4,A5,A6} such that the study of the discrete de Sitter isometries generating  the discrete symmetries of the Dirac theory was neglected so far.   For this reason we devote this paper to the discrete transformations of the covariant Dirac field on the de Sitter manifold.   

In special relativity the discrete transformations of the Dirac field $\psi$, written in our phases convention, are the parity, ${\cal P}\psi(t,{\bf x})=\gamma^0\psi(t,-{\bf x})$, charge conjugation ${\cal C}\psi(t,{\bf x})=i\gamma^2\psi^*(t,{\bf x})$, and Wigner's time reversal, ${\cal T}\psi(t,{\bf x})=-\gamma^1\gamma^3\psi^*(-t,{\bf x})$ \cite{Wig}  that closes the famous PCT theorem, ${\cal PCT}\psi=\gamma^5 \psi(-t,-{\bf x})$ \cite{BDR}.  As ${\cal T}$ is not related to an isometry, one uses often the transformation ${\cal CT}\psi(t,{\bf x})=\gamma^0\gamma^5\psi(-t,{\bf x})$ as an alternative time reversal \cite{KH}. The transformations ${\cal P}$, ${\cal CT}$ and ${\cal PCT}$ are produced by the {\em global} discrete $O(1,3)$ isometries $P$, $T$ and $PT$ transforming the Cartesian coordinates  $(t,{\bf x})$ into $(t,-{\bf x})$, $(-t,{\bf x})$ and respectively $(-t,-{\bf x})$.

For obtaining similar results for the Dirac field on the de Sitter manifold, $M$, we must focus on the discrete isometries of the $O(1,4)$ isometry group. On the other hand, we know that  only a portion of $M=M_+\cup M_-$ can play the role of physical space-time, either as an expanding ($M_+$) or as a collapsing universe ($M_-$). We have thus two types of discrete isometries, namely the {\em proper} ones that preserve the portion and the {\em improper} isometries which are changing these portions among themselves. In what follows we intend to study all these isometries even though the improper ones are of a mere mathematical interest. Thus we have to study three proper discrete isometries and four improper ones  extending the method of induced representations to the {\em local} discrete transformations of ${\rm Pin}O(1,3)$ we expect to meet here  as in this framework we can construct only three global discrete transformations, as in the flat case.  

The frames we need in our study are formed by a local chart and an orthogonal non-holonomic frame defined by tetrads. The discrete isometries of the de Sitter manifold were studied in a special coordonatization many time ago \cite{Mal} but unfortunately we cannot use these results as we need to work here with other types of coordinates.  We consider the conformal coordinates which are helpful in concrete calculations and the {\em physical} ones of Painlev\' e - Gullstrand type   \cite{Pan,Gul} which allow us to understand the physical meaning of the transformations studied here. Thus we have to work with conformal or physical frames in which we hope to obtain   a complete image about the discrete transformations of the covariant Dirac field produced by the de Sitter discrete isometries. 

We start in the next section introducing these frames in which we have to write down the Dirac equation after a brief review of the theory of covariant Dirac field presented is  Sec. 3.  Our study of the discrete de Sitter symmetries starts in Sec. 4 where we present the method of constructing the  induced representations that can be extended to the discrete de Sitter isometries we derive first in conformal frames separating the proper ones. For understanding their physical meaning we study in the next section the discrete symmetries generated by the proper isometries  in physical frames composed  with the charge conjugation. We find here a new {\em local } isometry reversing the time  helping us to define the analogous of the Wigner time reversal.  In this manner we obtain  the  discrete group of order 16 of the de Sitter expanding universe for which we derive the multiplication table.   Furthermore, for getting a global image about the discrete de Sitter isometries,  in Sec.  6  we come back to the conformal frames where we write down all of them including the improper ones.  Finally we present our concluding remarks.

\section{Frames}

Let us start with the  $(1+3)$-dimensional de Sitter manifold, $M$, defined as a hyperboloid   of radius $\omega_{H}^{-1}$ embedded in the five-dimensional flat space-time $M^5$ of coordinates $z^A$  (labeled by the indices $A,\,B,...= 0,1,2,3,4$) and metric $\eta^5={\rm diag}(1,-1,-1,-1,-1)$ \cite{BD},  denoting by $\omega_{H}$ the de Sitter-Hubble constant (frequency). A local chart of coordinates $x^{\mu}$ ($\mu,\nu,...=0,1,2,3$) can be introduced on $M$ giving the set of functions $z^A(x)$ which solve the hyperboloid equation, $\eta^5_{AB}z^A(x) z^B(x)=-\omega^{-2}_{H}$. 

The simplest local charts are the conformal ones, $\{x_c\}=\{t_c, {\bf x}_c\}$, with the {\em conformal} time $t_c$ and co-moving Cartesian space coordinates,  ${\bf x}_c=(x_c^1,x_c^2,x_c^3)$, defined by the functions
\begin{eqnarray}
	z_c^0(x_c)&=&-\frac{1}{2 \omega_{H}^2 t_c}
	\left[1- \omega_{H}^2\left(t_c^2-{{\bf x}_c\,}^2\right)\right]\,,\label{z0}\\
		z_c^i(x_c)&=&-\frac{x^i_c}{ \omega_{H} t_c}\,, \quad i,j,...=1,2,3\label{zi}\,,\\
	z_c^4(x_c)&=&-\frac{1}{2 \omega_{H}^2
		t_c}\left[1+ \omega_{H}^2\left(t_c^2-{{\bf x}_c\,}^2\right)\right]\,,\label{z4}
\end{eqnarray}
giving rise to the conformal-flat line element
\begin{equation}\label{FRW}
	ds^2=\eta^5_{AB}dz^A dz^B=\frac{1}{ \omega_{H}^2
		t_c^2}\,(dt_c^2-d{\bf x}_c\cdot d{\bf x}_c)\,.
\end{equation}
These charts, defined up to an isometry,   cover the expanding portion, $M_+$, for $t_c \in (-\infty,0)$ and ${\bf x}_c\in {\Bbb R}^3$ while the collapsing one, $M_-$,  is covered by  similar charts but with $t_c >0$ \cite{BD}. Note that the conformal coordinates for FLRW space-times were introduced by Lema\^ itre \cite{Lemco}.

However, the coordinates with the obvious physical meaning are those of Painlev\' e - Gullstrand  type  \cite{Pan,Gul} defining the {\em physical} local chart $\{x\}=\{ t,{\bf x}\}$. These coordinates are  the {\em cosmic} time $t$ and the physical space coordinates ${\bf x}=(x^1,x^2,x^3)$ that can be  introduced on the expanding portion $M_+$ by substituting $x_c=\chi(x)$ where
\begin{equation}\label{subs}
	t_c=\chi^0(x)=-\frac{1}{\omega_{H}}e^{-\omega_{H} t}\,,  \quad {x}^i_c=\chi^i(x)=x^ie^{-\omega_{H} t}\,,
\end{equation}
and obtaining the new line element
\begin{equation}\label{Pan}
	ds^2=\left(1-\omega_{H}^2 {{\bf x}}^2\right)dt^2+2\omega_{H} {\bf x}\cdot d{\bf x}\,dt -d{\bf x}\cdot d{\bf x}\,. 
\end{equation}
Similar coordinates can be defined on the collapsing portion $M_-$ following the same procedure but with $\omega_{H}\to -\omega_{H}$.  

For writing down the Dirac equation  we need to set the tetrad gauge giving  the vector fields $e_{\hat\alpha}=e_{\hat\alpha}^{\mu}\partial_{\mu}$ defining the local orthogonal frames,  and the 1-forms $\omega^{\hat\alpha}=\hat e_{\mu}^{\hat\alpha}dx^{\mu}$ of the dual co-frames (labelled by the local indices, $\hat\mu,\hat\nu,...=0,1,2,3$).    Here we restrict ourselves to the Cartesian diagonal tetrad gauge defined by the vector fields  
\begin{eqnarray}
	e_{0}&=&-\omega_{H}\,t_c\partial_{t_c}=\partial_t+\omega_{H}\,x^i\partial_{x^i}\,,\nonumber \\
	e_{i}&=&-\omega_{H}\,t_c\partial_{x^i_c}=\partial_{x^i}\,,\label{e}
\end{eqnarray}
and the corresponding dual 1-forms 
\begin{eqnarray}
	\omega^{0} &=&-\frac{1}{\omega_{H}\,t_c}dt_c=dt\,,\nonumber \\
	\omega^{i} &=& -\frac{1}{\omega_{H}\,t_c}dx_c^i=d{x}^i-\omega_{H}\,{x}^i dt\,,\label{o}
\end{eqnarray}
which preserve the global $SO(3)$ symmetry  allowing us to use systematically the  $SO(3)$ vectors.  We obtain thus the conformal  frames  $\{x_c;e\}$ and the physical ones $\{x;e\}$  formed by a  local chart, $\{x_c\}$ or $\{x\}$, and a local orthogonal frame and co-frame given by the tetrads $e$ and $\hat e$, as defined by Eqs. (\ref{e}) and (\ref{o}) respectively. We remind the reader that each physical frame  is the proper frames of an observer staying at rest in origin measuring the events inside the event horizon, for $|{\bf x}|<\omega_H^{-1}$.
Therefore, the physical interpretation must be done in physical frames while the conformal ones are  precious tools in our investigations.

These frames, say the physical ones,  can be transformed, $\{x;e\}\to \{x';e'\}$, with the help of  diffeomorphisms, $x\to x'=\phi(x)$, transforming the coordinates, and by using  local transformations $\Lambda(x)\in L^{\uparrow}_{+}$, for changing the tetrad gauge as 
\begin{eqnarray}
	e_{\hat\alpha}(x)&\to & e_{\hat\alpha}'(x)=\Lambda_{\hat\alpha\,\cdot}^{\cdot\,\hat\beta}(x)\, e_{\hat\beta}(x)\,,\label{gauge1}\\
	{\omega}^{\hat\alpha}(x)&\to &{ \omega}^{\prime\,\hat\alpha}(x)=\Lambda^{\hat\alpha\,\cdot}_{\cdot\,\hat\beta}(x) \,\omega_{\hat\beta}(x)\,.\label{gauge2}
\end{eqnarray}
In general relativity any theory of fields with spin  must be gauge invariant in the sense that the above gauge transformations have to do not affect the physical meaning of the theory. However, as here we intend to study discrete transformations we add them to the gauge group which will be denoted from now by $G=O(1,3)$. 

\section{Covariant Dirac field}

In a frame $\{x;e\}$ of  $M_+$ the tetrad-gauge invariant action of the Dirac field $\psi : M_+\to {\cal V}_D$, of mass $m$, minimally coupled to the background gravity, reads,  
\begin{equation}\label{action}
	{\cal S}[e,\psi]=\int\, d^{4}x\sqrt{g}\left\{
	\frac{i}{2}[\overline{\psi}\gamma^{\hat\alpha}\nabla_{\hat\alpha}\psi-
	(\overline{\nabla_{\hat\alpha}\psi})\gamma^{\hat\alpha}\psi] -
	m\overline{\psi}\psi\right\}
\end{equation}
where  $\bar{\psi}=\psi^+\gamma^0$ is the Dirac adjoint of $\psi$ and  $g=|\det(g_{\mu\nu})|$. 
The field $\psi$ takes values in the space of Dirac spinors ${\cal V}_D$ which carries the Dirac representation $\rho_D= (\frac{1}{2},0)\oplus (0,\frac{1}{2})$ of the $SL(2,{\Bbb C})$ group. In this representation one can define the Dirac matrices $\gamma^{\hat\alpha}$ (with local indices) which are Dirac self-adjoint, $\overline{\gamma^{\hat\mu}}=\gamma^0{\gamma^{\hat\mu}}^+\gamma^0=\gamma^{\hat\mu}$, and satisfy the anti-commutation rules $\{ \gamma^{\hat\alpha},\, \gamma^{\hat\beta} \}=2\eta^{\hat\alpha \hat\beta}$. These matrices may extend the $sl(2,{\Bbb C})$ Lie algebra to a  $su(2,2)$ one such that $\rho_D$ is in fact much larger than a representation of the $SL(2,{\Bbb C})$ group. 

The gauge invariant theory releases on the covariant derivatives defined as
\begin{equation}\label{cov}
	\nabla_{\hat\alpha}=e^{\mu}_{\hat\alpha}\nabla_{\mu}= e_{\hat\alpha}+\frac{i}{2}\hat\Gamma^{\hat\gamma}_{\hat\alpha
		\hat\beta}s^{\hat\beta\,
		\cdot}_{\cdot\, \hat\gamma}\,,\quad 	s^{\hat\alpha \hat\beta}=\frac{i}{4}\,[\gamma^{\hat\alpha}, 	\gamma^{\hat\beta} ]\,,
\end{equation}
where $s^{\hat\alpha \hat\beta}$ are the $SL(2,{\Bbb C})$ generators of $\rho_D$ while 
$\hat\Gamma^{\hat\sigma}_{\hat\mu \hat\nu}=e_{\hat\mu}^{\alpha} e_{\hat\nu}^{\beta}
(\hat e_{\gamma}^{\hat\sigma}\Gamma^{\gamma}_{\alpha \beta} -\hat e^{\hat\sigma}_{\beta, \alpha})$ are the connection components in local frames (known as the spin connections and denoted often by $\Omega$) expressed in terms of tetrads and Christoffel symbols,   $\Gamma^{\gamma}_{\alpha \beta}$. Thanks to these derivatives, the Dirac equation
\begin{equation}\label{EEE}
	\left(  i\gamma^{\hat\alpha}\nabla_{\hat\alpha}-m\right)\psi(x)=0\,,
\end{equation}
is {\em gauge invariant} in the sense that this does not change its form when we perform the  transformation 
\begin{equation}
	\psi(x)\to  \psi'(x)=  \lambda(x)\psi(x)\,,\label{Gauge}
\end{equation}
simultaneously with the gauge transformation (\ref{gauge1}) and (\ref{gauge2}) with $\Lambda(x)$ depending on $\lambda(x)$ through the canonical homomorphism \cite{WKT}.  We remind the reader that in the covariant parametrization with skew-symmetric real valued parameters, $\hat\omega_{\hat\alpha\hat\beta}=-\hat\omega_{\hat\beta\hat\alpha}$, the $SL(2,{\Bbb C})$ transformations 
\begin{equation}
	\lambda(\hat\omega)=\exp\left(-\frac{i}{2}\,\hat\omega_{\hat\alpha\hat\beta}s^{\hat\alpha\hat\beta}\right)\in\rho_D[SL(2,{\Bbb C})]\,.
\end{equation}
correspond through the canonical homomorphism to the transformation matrices
$\Lambda(\hat\omega)\in L^{\uparrow}_{+}$ having the matrix elements $\Lambda(\hat\omega)^{\hat\alpha\,\cdot}_{\cdot \,\hat\beta}=\delta^{\hat\alpha}_{\hat\beta}+\hat\omega^{\hat\alpha\,\cdot}_{\cdot \,\hat\beta}+...$. Note that this property is encapsulated in the identity 
\begin{equation}
\lambda^{-1}(\hat\omega)\gamma^{\hat\alpha}\lambda(\hat\omega)	=\Lambda(\hat\omega)^{\hat\alpha\,\cdot}_{\cdot \,\hat\beta}\gamma^{\hat\beta}\,,
\end{equation}
that holds for any $\lambda(\hat\omega)\in \rho_D[SL(2,{\Bbb C})]$. Moreover, we assume that this identity can be used even for relating the discrete transformations of $\rho_D$,  constructed with the help of the $\gamma$-matrices, to those of $G$. For shortening the notation we denote by $\tilde G={\rm Pin}\,O(1,3)$ the group associated to the gauge group $G=O(1,3)$ through canonical hamomorphism, understanding that $\rho_D$ includes a representation of $\tilde G$. 

According to this general formalism the Dirac equation in the conformal frame  $\{x_c;e\}$ of $M_+$ reads \cite{CD1}
\begin{equation}\label{ED1}
	\left[-i\omega_{H}t_c\left(\gamma^0\partial_{t_c}+\gamma^i\partial_{x^i_c}\right)
	+\frac{3i}{2}\omega_H\gamma^{0}-m\right]\psi(t,{\bf x})=0\,,
\end{equation}
keeping the same form on $M_-$ where $t_c>0$. In contrast, the Dirac equation in the physical frames $\{x;e\}$ of $M_+$, \cite{CSchr} 
\begin{equation}\label{ED2}
	\left[i\gamma^0\partial_{t}+i{\gamma^i}{\partial_{x^i}} -m
	+i\gamma^{0}\omega_{H}
	\left({x}^i{\partial}_{x^i}+\frac{3}{2}\right)\right]\psi(t,{\bf x})=0\,,
\end{equation}
is different from that on $M_-$ which has a similar form but with $\omega_H\to-\omega_H$ changing thus the sign of the last term due to the de Sitter gravity. Both these equations can be solved analytically. For example, in conformal frame the general solution   may be written as a mode integral \cite{CD1}, 
\begin{eqnarray}
	\psi({x}_c\,)= \int d^{3}p
	\sum_{\sigma}[U_{{\bf p},\sigma}(x_c){\alpha}({\bf p},\sigma)
	+V_{{\bf p},\sigma}(x_c){\beta}^{*}({\bf p},\sigma)]\,,\label{p3}
\end{eqnarray}
in terms of wave functions in momentum representation of particle, $\alpha$, and antiparticle, $\beta$.  The mode spinors $U_{{\bf p},\sigma}$  and  $V_{{\bf p},\sigma}$, of positive and respectively negative frequencies, are plane waves solutions of the Dirac equation depending on the conserved momentum ${\bf p}$ and an arbitrary polarization $\sigma$. These spinors  form an orthonormal  basis being related  through the usual charge conjugation, 
\begin{equation}\label{chc}
	V_{{\bf p},\sigma}({x}_c)=i\gamma^2\left[{U}_{{\bf p},\sigma}({x}_c)\right]^* \,, 
\end{equation}
and satisfying orthogonality relations and a completeness condition with respect to a suitable relativistic scalar product \cite{CD1}. Similar properties hold in physical frames such that we may consider the charge conjugation
\begin{equation}\label{C}
	{\cal C}\psi =i\gamma^2 \psi^*\,,
\end{equation} 
on the whole manifold $M$ where this has the same properties as in the flat case  being    independent on geometry.

\section{Continuous and discrete isometries}

The de Sitter manifold $M=M_+\cup M_-$ has the isometries generated by the group $O(1,4)$ that leaves the metric $\eta^5$ invariant. In general, these isometries, $x\to x'=\phi_{\frak g} (x)$, are defined for any ${\frak g}\in O(1,4)$ such that $z[\phi_{\frak g}(x)]={\frak g}z(x)$ forming  the isometry group $I(M)\sim O(1,4)$ with respect to the composition rule $\phi_{\frak g}\circ \phi_{{\frak g}'}=\phi_{\frak{gg}'}$, $\forall \frak{g,g}'\,\in I(M)$. The identity element of  $I(M)$ is the  identity function ${\rm id}=\phi_{\frak e}$, coresponding to the identity element  ${\frak e}$ of the group $O(1,4)$. Therefore, the calculation rules $\phi\circ {\rm id}={\rm id}\circ\phi=\phi$ and $\phi\circ \phi^{-1}=\phi^{-1}\circ\phi ={\rm id}$ hold for all $\phi\in I(M)$.  

The problem here is that the isometries may change the tetrad  gauge modifying thus  the form of the Dirac equation.  For assuring the invariance under isometries of the Dirac or other field equations  one of us  introduced the combined transformations $(\lambda_{\frak g},\phi_{\frak g})$,  formed by the mappings $\lambda_{\frak g} : x\to \lambda_{\frak g}(x)\in\rho_D$ and $\phi_{\frak g} : x\to \phi_{\frak g}(x)$,  which have to correct in each point the positions of the local frames after an isometry. In other words, these transformations must preserve simultaneously  the metric  and the tetrad gauge as  $\omega(x')=\Lambda[\lambda_{\frak g}(x)]\omega(x)$ having the form    \cite{ES},
\begin{equation}\label{Axx}
	\Lambda^{\hat\alpha\,\cdot}_{\cdot\,\hat\beta}[\lambda_{\frak g}(x)]= \hat
	e_{\mu}^{\hat\alpha}[\phi_{\frak g}(x)]\frac{\partial
		\phi^{\mu}_{\frak g}(x)} {\partial x^{\nu}}\,e^{\nu}_{\hat\beta}(x)\,,
\end{equation}
which define the mapping $\lambda_{\frak g}$ assuming, in addition,  that  
\begin{equation}\label{cond}
	\lambda_{{\frak g}={\frak e}}(x) =1\in \rho_D\,. 	
\end{equation}
The resulted combined transformations $(\lambda_{\frak g},\phi_{\frak g})$ preserve the gauge,  $e'=e$ and $\omega'=\omega$,  transforming the  Dirac field according to the covariant representation  $\mathsf{ T} : (\lambda_{\frak g},\phi_{\frak g})\to \mathsf{ T}_{\frak g}$   whose operators act as
\begin{equation}\label{Tx}
\left(\mathsf{T}_{\frak g}\psi\right)[\phi_{\frak g}(x)]=\lambda(x)\psi(x)~~\Rightarrow~~	\mathsf{ T}_{\frak g}\psi=(\lambda_{\frak g} \psi) \circ \phi_{\frak g}^{-1}\,.
\end{equation}
Thus the frame transformation $(\lambda_{\frak g},\phi_{\frak g}) : (x;e)\to \left( \phi_{\frak g}(x);\Lambda(\lambda_{\frak g})\, e \right)$ and the covariant one (\ref{Tx}) leave the Dirac equation invariant preserving the physical meaning of the entire theory.   In the de Sitter frames we consider here, $\{x_c;e\}$ and $\{x;e\}$, we denote the combined transformations by $(\lambda_{\frak g}^c, \phi_{\frak g}^c)$ and respectively  $(\lambda_{\frak g}, \phi_{\frak g})$ bearing in mind that these are related as $\phi_{\frak g}=\phi_{\frak g}^c\circ \chi$  and $\lambda_{\frak g}=\lambda_{\frak g}^c\circ \chi$ where the function $\chi$ is defined in Eq. (\ref{subs}). 

Mathematically speaking, we must specify that the mappings $\lambda:M\to \tilde G$ are sections on the principal fiber bundle $M\times \tilde G$ while the mapping associated through canonical homomorphism, $\Lambda(\lambda):M\to G$ are sections on $M\times G$. These mappings can be composed as $\lambda_{{\frak g}'}\times \lambda_{\frak g}$ such that ($\lambda_{{\frak g}'}\times \lambda_{\frak g})(x)=\lambda_{{\frak g}'}(x) \lambda_{\frak g}(x)$ and similarly for the mappings $\Lambda$. Moreover, observing that $(\lambda_{{\frak g}'}\circ \phi_{\frak g})\times \lambda_{\frak g}=\lambda_{{\frak g}'{\frak g}}$ we may conclude that the pairs $(\lambda_{\frak g},\phi_{\frak g})$ form a well-defined Lie group with respect to the new operation  \cite{ES}
\begin{equation}\label{mult}
	(\lambda_{{\frak g}'},\phi_{{\frak g}'})*(\lambda_{\frak g},\phi_{\frak g})=((\lambda_{{\frak g}'}\circ \phi_{\frak g})\times \lambda_{\frak g}, \phi_{{\frak g}'}\circ \phi_{\frak g})
	=	(\lambda_{{\frak g}'{\frak g}},\phi_{{\frak g}'{\frak g}})\,.
\end{equation}  
 This group can be seen as a representation of the universal covering group of $I(M)$. Therefore, we may say that  the covariant representations defined in Ref. \cite{ES} transfer the Lorentz covariance from special  to general relativity. 

The discrete isometries are generated by the discrete transformations ${\frak d}$ of the group $O(1,4)$  that  satisfy ${\frak d}{\frak d}={\frak e}$. Each discrete transformation ${\frak d}$ defines the combined transformation $(\lambda_{\frak d}, \phi_{\frak d})$ whose mappings $\lambda_{\frak d} :x \to \lambda_{\frak d}(x)$ give rise to local discrete transformations $\lambda_{\frak d}(x)\in \rho_D$ which must satisfy $\lambda_{\frak d}(x) \lambda_{\frak d}(x)=\pm 1\in\rho_D$. However, the corresponding transformations through the canonical homomorphism have to obey $\Lambda(\lambda_{\frak d})\Lambda(\lambda_{\frak d})=I$ where $I$ is the identity of  the  group $G$. In general, the discrete transformations $\lambda_{\frak d}(x)$ are {\em  local} depending on $x$ but there are {\em global} discrete transformations which reduce to simple matrices independent on $x$. For example, the matrices $\gamma^0$, $\gamma^0\gamma^5$ and $\gamma^5$ of $\rho_D$ are associated through the canonical homomorphism to the global discrete transformations,
\begin{eqnarray}
\Lambda(\gamma^0)&=&P={\rm diag}(1,-1,-1,-1)\,,\\	
\Lambda(\gamma^0\gamma^5)&=&T={\rm diag}(-1,1,1,1)\,,\label{TT}\\
\Lambda(\gamma^5)&=&PT=-I,,
\end{eqnarray} 
giving rice to the subsets of the Lorentz group  $G$ \cite{WKT}.    

The de Sitter discrete isometries $\phi^c_{\frak d}$ in  the conformal frames  given by the functions (\ref{z0}-\ref{z4})  solve the equations $z_c[\phi^c_{\frak d}(x_c)]={\frak d}z_c(x_c)$. The simplest discrete transformations on $M^5$ are the mirror reflections, denoted by $[A]$, which change the sign of a single coordinate, $z_c^A\to -z_c^A$, giving rise to the discrete isometries $x_c\to x_c'=\phi^c_{[A]}(x_c)$.  A rapid inspection indicates that there are two interesting local isometries  produced by $\phi^c_{[0]}$ as,
\begin{equation}\label{Pi0}
	t_c'(x_c)=\phi_{[0]}^{c\,0}(x_c)=\frac{t_c}{\omega^2 (t_c^2-{\bf x}_c^2)}\,, \quad
	x_c^{\prime\, i}(x_c)=\phi_{[0]}^{c\,i}(x_c)=\frac{x^i_c}{\omega^2
		(t_c^2-{\bf x}_c^2)}\,,
\end{equation}
and by $\phi^c_{[4]}$ which gives
\begin{equation}\label{Pi4}
	\phi_{[4]}^{c\,0}(x_c)=-	t_c'(x_c)\,, \qquad
\phi_{[4]}^{c\,i}(x_c)=-x_c^{\prime\, i}(x_c)\,.
\end{equation}
The space isometries, $\phi^c_{[i]}$, are simple mirror transformations of the space co-moving coordinates,  $x_c^i\to -x_c^i$ such that the space parity can be defined globally as  $\phi^c_{\frak p}=\phi^c_{[1]}\circ\phi^c_{[2]}\circ\phi^c_{[3]}$.  Moreover, we can construct the isometries  $\phi^c_{{\frak p}[0]}=\phi^c_{\frak p}\circ\phi^c_{[0]} $, $\phi^c_{{\frak p}[4]}=\phi^c_{\frak p}\circ\phi^c_{[4]} $  and  $\phi^c_{[0][4]}=\phi^c_{[0]}\circ\phi^c_{[4]}$, the last one changing  the signs of all the conformal coordinates,  $x_c^{\mu}\to- x_c^{\mu}$.  The antipodal transformation, ${\frak a} : z_c\to -z_c$, gives rise to the isometry $\phi^c_{\frak a}=\phi^c_{\frak p}\circ\phi^c_{[0][4]}$ that reverses the conformal time on $M$, $t_c\to -t_c$. 

Hereby we see that the set  of discrete isometries ${\frak D}={\frak D}_+\cup {\frak D}_-$ can be split in two subsets, ${\frak D}_+=\{\phi_{\frak d} | \phi_{\frak d} : M_{\pm}\to M_{\pm}\}$  of {\em proper} discrete isometries which preserve the portions of $M$ and ${\frak D}_-=\{\phi_{\frak d} | \phi_{\frak d} : M_{\pm}\to M_{\mp}\}$ of  {\em improper} discrete  isometries that change these portions among themselves.  As  the physical measurements can be performed only inside the light-cone where $|t_c|>|{\bf x}_c|$ we deduce that the isometry (\ref{Pi0})  preserve the sign of conformal time while the isometry  (\ref{Pi4}) changes it. Therefore, we may conclude that 
\begin{eqnarray}
\phi_{\frak e}={\rm id},\,\phi^c_{\frak p},\,\phi^c_{[0]},\,\phi^c_{{\frak p}[0]}&\in& {\frak D}_+\label{Dp}\\
\phi^c_{\frak a},\,\phi^c_{[4]},\,\phi^c_{[0][4]},\,\phi^c_{{\frak p}[4]}&\in& {\frak D}_-\label{Dm}
\end{eqnarray}
and similarly for the isometries $\phi_{\frak d}$ of the physical frames. For understanding the physical meaning of these isometries we discuss first the proper discrete isometries  in physical frames.  

\section{Discrete symmetries in physical frames}

 Let us consider now the physical frames of the expanding portion $M_+$ known as the de Sitter expanding universe. The physical coordinates on $M_+$ can be introduced by  the new functions $z=z_c\circ \chi$ that read
\begin{eqnarray}
	z^0(x)&=&\frac{1}{2 \omega_{H}}
	\left[e^{\omega_H t}-e^{-\omega_H t} \left(1-\omega_{H}^2{{\bf x}}^2\right)\right]\,,\label{z10}\\
	z^i(x)&=&x^i\label{z1i}\,,\\
	z^4(x)&=&\frac{1}{2 \omega_{H}}
	\left[e^{\omega_H t}+e^{-\omega_H t} \left(1-\omega_{H}^2{{\bf x}}^2\right)\right]\,.\label{z14}
\end{eqnarray}
Similar functions that define the physical coordinates on $M_-$ can be obtained changing $\omega_H\to -\omega_H$ in Eqs. (\ref{z10}-\ref{z14}).

We observe first that the trivial isometry id is needed for closing a discrete group. Denoting $\mathsf{T}_{\frak e}={\cal I}$ we relax the condition (\ref{cond}) for recovering the kernel $\mathbb{Z}_2=\{{\cal I},\, -{\cal I}\}$ of the canonical homomorphism we need when we study discrete transformations.  The simplest non-trivial discrete isometry is  the parity  ${\frak p}$ which is global on $M^5$  such that $\phi^0_{\frak p}(x)=t$, $\phi^i_{\frak p}(x)=-x^i$ and  $\lambda_{\frak p}=\gamma^0$. Denoting then $\mathsf{T}_{\frak p}={\cal P}$ we obtain the parity transformation of the Dirac field,
\begin{equation}\label{parity}
	{\cal P}\psi(t,{\bf x})=\gamma^0 \psi(t,-{\bf x})\,,
\end{equation}	
acting just as in the flat case but transforming, in addition, the tetrads fields with $\Lambda_{\frak p}=P$.     

\begin{table}
	\begin{tabular}{ccccccc}
		$~{\frak d}$~~&~~$T_{\frak d}$~~&~~$\phi^{0}_{\frak d}(x)~~$&~~$\phi^{i}_{\frak d}(x)$~~&$\lambda_{\frak d}(x)$&$\Lambda_{\frak d}(x)$&~~Eq.~~\\	
		&&&&&&\\
		~~${\frak p}$~~&${\cal P}$&$t$&$-{x}^{i}$&$\gamma^0$&$P$&(\ref{parity})\\
		$[0]$&${\cal B}$&$t'(t, |{\bf x}|)$&${x}^i$&$\gamma^0\gamma^5\beta({\bf x})$~~&$TB({\bf x})$& (\ref{barity})\\
		~${\frak p}[0]$~&~~~${\cal PB}$~~~&$t'(t, |{\bf x}|)$&$-{x}^{i}$&$\gamma^5\beta(-{\bf x})$&$-B(-{\bf x})$~~~&(\ref{carity}) 		
	\end{tabular}
	\caption{Transformations induced by the non-trivial proper discrete  isometries in physical frames of the de Sitter expanding universe, $M_+$. The function $t'(t, |{\bf x}|)$ is defined in Eq. (\ref{loct}) while the matrices $B({\bf x})$ and $\beta({\bf x})$ are given by Eqs. (\ref{Bx}) and respectively (\ref{betax}). }
\end{table}	

Furthermore, we consider the isometry $x'=\phi_{[0]}(x)$ generated by the mirror transformation $[0]$ which changes $z^0\to -z^0$. Remarkably,  this is a {\em local time reversal},  	
\begin{eqnarray}
t'(t, |{\bf x}|)&=&\phi_{[0]}^0(x)=-t	+\frac{1}{\omega_H}\ln\left(1-\omega_H^2 {\bf x}^2\right)\,,\label{loct}\\
x^{\prime\,i}&=&\phi_{[0]}^i(x)=x^i\,,
\end{eqnarray}
giving the new local time  $t'(x)\le -t$  but without affecting the space coordinates. The problem now is to find the associated matrices $\lambda_{[0]}$ and $\Lambda_{[0]}\equiv\Lambda(\lambda_{[0]})$ transforming the Dirac and respectively the tetrad fields. We first apply the definition (\ref{Axx}) for deriving the local  matrix $\Lambda_{[0]}(x)$ finding that this is time-independent having the matrix elements
\begin{eqnarray}
\left<\Lambda_{[0]}({\bf x})\right>^{0\,\cdot}_{\cdot\,0}&=&-\frac{1+\omega^2_H{\bf x}^2}{1-\omega^2_H{\bf x}^2}<0\,,	\nonumber\\ 
\left<\Lambda_{[0]}({\bf x})\right>^{0\,\cdot}_{\cdot\,i}&=&-\left<\Lambda_{[0]}(x)\right>^{i\,\cdot}_{\cdot\,0}=-\frac{2\omega_H x^i}{1-\omega^2_H{\bf x}^2}\,,\\
\left<\Lambda_{[0]}({\bf x})\right>^{i\,\cdot}_{\cdot\,j}&=&\delta^i_j+\frac{2\omega^2_H x^i x^j}{1-\omega^2_H{\bf x}^2}\,.\nonumber
\end{eqnarray}
Moreover, we find that $\det[\Lambda_{[0]}{(\bf x})]=-1$ which means that $\Lambda_{[0]} \in TL^{\uparrow}_{+}$. Therefore, we can use the factorization $\Lambda_{[0]}({\bf x})=T B({\bf x})$ where now   $B({\bf x})\in L^{\uparrow}_{+}$. It is not difficult to verify that $B({\bf x})$  is a Lorentz boost that can be parametrized as  
\begin{equation}\label{Bx}
	B({\bf x})=\exp\left(-iK_i\,\frac{x^i}{|{\bf x}|}\, \alpha( |{\bf x}|)\right) \,,\quad \alpha( |{\bf x}|)={\rm tanh}^{-1}\left(\frac{2 \omega_H  |{\bf x}|}{1+\omega^2_H{\bf x}^2}\right)\,,
\end{equation}
using the boost generators $K_i$ of the $L^{\uparrow}_{+}$ group (having the non-vanishing components $\left<K_i \right>^{j}_{0} =\left<K_i \right>_{j}^{0}=i\delta_{ij}$ \cite{WKT}). Note that these boost matrices satisfy $B({\bf x})^{-1}=B(-{\bf x})$ and $B({\bf x}=0)=I$. In addition, we can verify that 
\begin{equation}
	\Lambda_{[0]}({\bf x})[\Lambda_{[0]}({\bf x})=I~ ~\Longleftrightarrow~~ B({\bf x})TB({\bf x})=T\,,
\end{equation}
as the pseudo-orthogonal boosts $B({\bf x})$ are symmetric, $B({\bf x})^T=B({\bf x})$.

The above parametrization helps us to find the local matrices $\lambda_{[0]}({\bf x})$ which transform the Dirac field. According to Eq. (\ref{TT}) this can be written as
\begin{equation}
\lambda_{[0]}({\bf x})=\gamma^0\gamma^5 \beta({\bf x})\,,\quad  \,,	
\end{equation}  
where  the matrix 
\begin{equation}\label{betax}
\beta({\bf x})=\exp\left(-is_{i0}\,\frac{x^i}{|{\bf x}|}\, \alpha( |{\bf x}|)\right)=\frac{1+\omega_H \gamma^0\gamma^i x^i}{\sqrt{1-\omega^2_H{\bf x}^2}}\in \rho_D\,,	
\end{equation}
is calculated by using the boost generators  $s_{i0} \in\rho_D$  defined in Eq. (\ref{cov}). This has the obvious properties, $\beta({\bf x})^{-1}=\beta(-{\bf x})$ and $\gamma^0\beta({\bf x})=\beta(-{\bf x})\gamma^0$ which guarantee that $\lambda_{[0]}({\bf x})\lambda_{[0]}({\bf x})=-1\in\rho_D$. Moreover, we verify that the matrices $\lambda_{[0]}({\bf x})$ and $\Lambda_{[0]}({\bf x})$ are related through the canonical homomorphism satisfying the identities
\begin{equation}
	\lambda_{[0]}({\bf x})^{-1}\gamma^{\hat\alpha}	\lambda_{[0]}({\bf x})=\left<\Lambda_{[0]}({\bf x})\right>^{\hat\alpha\,\cdot}_{\cdot\,\hat\beta}\gamma^{\hat\beta}\,,
\end{equation}
which assure the invariance of the Dirac equation (\ref{ED2}) under this isometry.

Collecting then the above results and denoting $\mathsf{T}_{[0]}={\cal B}$ we define the discrete transformation of the Dirac field 
\begin{equation}\label{barity}
	{\cal B}\psi (t,{\bf x})=	\lambda_{[0]}({\bf x})\psi(t'(x), {\bf x})=\gamma^0\gamma^5\beta({\bf x})\psi(t'(t, |{\bf x}|)), {\bf x})\,,
\end{equation}
where $t'(x)$ is given by Eq. (\ref{loct}).  The above results help us to write the transformation of the Dirac field associated to the isometry $\phi^c_{{\frak p}[0]}$  as 
\begin{equation}\label{carity}
{\cal PB}\psi(t,{\bf x})=\gamma^5 \beta(-{\bf x})\psi(t'(t, |{\bf x}|), -{\bf x} )\,,	
\end{equation} 
while the tetrads have to be transformed by $\Lambda_{{\frak p}[0]}=PTB(-{\bf x})=-B(-{\bf x})$. Obviously, the transformations ${\cal B}$ and ${\cal PB}$  take over the role of ${\cal CT}$ and respectively ${\cal PCT}$ ones of the Dirac theory in Minkowski space-time.

\begin{table}
	\begin{tabular}{rrrrrrrr}
		~~~~~${\cal I }$~~&~~~~~${\cal C}$~~&~~~~~${\cal P}$~~&~~~~${\cal PC }$~~&~~~~${\cal  T}_+$&~~~~${\cal P}{\cal T}_+$&~~~~~${\cal C}{\cal T}_+$&~~~~${\cal PC}{\cal T}_+$\\
		${\cal C}$~~&${\cal I}$~~&$-{\cal PC }$~~&$-{\cal P}$~~&${\cal C}{\cal T}_+$&$-{\cal PC}{\cal T}_+$&${\cal T}_+$&$-{\cal P}{\cal T}_+$\\
		${\cal P}$~~&${\cal PC}$~~&${\cal I}$~~&${\cal C}$~~&${\cal P}{\cal T}_+$&${\cal T}_+$&${\cal PC }{\cal T}_+$&${\cal  C}{\cal T}_+$\\
		${\cal PC}$~~&${\cal  P}$~~&$-{\cal  C}$~~&$-{\cal I}$~~&${\cal PC}{\cal T}_+$&$-{\cal C}{\cal T}_+$&${\cal P}{\cal T}_+$&$-{\cal T}_+$\\
		${\cal T}_+$&${\cal C}{\cal T}_+$&${\cal P}{\cal T}_+$&${\cal PC}{\cal T}_+$&$-{\cal I}~~$&$-{\cal P }$~~&$-{\cal C}$~~&$-{\cal PC}$~~\\	
		${\cal P}{\cal T}_+$&${\cal PC}{\cal T}_+$&${\cal T}_+$&${\cal C}{\cal T}_+$&$-{\cal P}$~~&$-{\cal I}$~~&$-{\cal PC}$~~&$-{\cal C}$~~\\
		${\cal C}{\cal T}_+$&${\cal T}_+$&$-{\cal PC}{\cal T}_+$&$-{\cal P}{\cal T}_+$&$-{\cal C}$~~&${\cal PC }$~~&$-{\cal I}$~~&$-{\cal P }$~~\\
		${\cal PC}{\cal T}_+$&${\cal P}{\cal T}_+$&$-{\cal C}{\cal T}_+$&$-{\cal T}_+$&$-{\cal PC}$~~&${\cal C}$~~&$-{\cal P}$~~&$-{\cal I}$~~
	\end{tabular}
	\caption{Multiplication table of the subset ${\cal G}/{\Bbb Z}_2$ of the group ${\cal G}$ of the discrete transformations of the covariant Dirac field in the de Sitter expanding universe $M_+$.}
\end{table} 
We see thus that the Dirac theory on $M_+$  has  three non-trivial proper discrete isometries, ${\cal P},\, { \cal B}$ and ${\cal PB}=-{\cal BP}$, that can be composed at any time with the charge conjugation ${\cal C}$ which is independent on geometry. Under such circumstances we may define  of the Wigner time reversal on $M_+$ as  ${\cal T}_+={\cal C}{{\cal B}}=\cal{BC}$, obtaining again a local transformation which in our phase convention reads, 
\begin{eqnarray}
{\cal T}_+\psi(t,{\bf x})&=&i\gamma^2\left( {\cal B}\psi(t,{\bf x})\right)^*=-\beta(-{\bf x})\gamma^1\gamma^3\psi^*(t'(t, |{\bf x}|),{\bf x})\,,	\label{Wtr}
\end{eqnarray} 
as the matrix $\beta({\bf x})$ satisfies $\gamma^1\gamma^3\beta^*({\bf x})=\beta(-{\bf x})\gamma^1\gamma^3$. Another transformation can be defined as
\begin{equation}\label{PWtr}
	{\cal P}{\cal T}_+\psi(t,{\bf x})=-\beta({\bf x})\gamma^0\gamma^1\gamma^3\psi^*(t'(t, |{\bf x}|),{-\bf x})\,,
\end{equation}
observing that ${\cal P}{\cal T}_+={\cal T}_+{\cal P}$. Note that under these last two transformations  the tetrad fields have to transform according to the transformations $\Lambda_{[0]}$ and respectively $\Lambda_{{\frak p}[0]}$  as ${\cal C}$ does not produce geometric effects. 

Finally, we substitute ${\cal B}={\cal C}{\cal T}_+$ obtaining the standard components   
 \begin{equation}
{\cal G}_+=\{ \pm{\cal I},\,\pm{\cal C},\,\pm{\cal P},\,\pm{\cal PC},\,\pm{\cal T}_+,\,\pm{\cal P}{\cal T}_+,\,\pm{\cal C}{\cal T}_+,\,\pm {\cal PC}{\cal T_+}\} 	
 \end{equation}
of the discrete group  of Dirac's theory on the de Sitter expanding universe, $M_+$. This group is of the order 16 having the multiplication table given in Tab. 2 which is similar to  that of the Dirac theory in Minkowski space-time.   In this group we meet the usual parity ${\cal P}$ and charge conjugation ${\cal C}$ which are global while all the transformations involving the Wigner time reversal ${\cal T}_+$ are local, depending on the matrix $\beta({\bf x})$.  Similar results can be obtained for the collapsing portion where we meet a similar discrete group ${\cal G}_-$ but with the Wigner time reversal ${\cal T}_-$ that can be obtained from ${\cal T}_+$ substituting $\omega_H\to-\omega_H$.  An important opportunity of this approach is the correct flat limit when $\omega_H\to 0$. Indeed,  in Eq. (\ref{loct}) we have
\begin{equation}
	\lim_{\omega_H\to 0}\frac{1}{\omega_H}\ln(1-\omega_H^2{\bf x}^2)=0 ~~\Rightarrow~~ 	\lim_{\omega_H\to 0}{\cal T}_+=\lim_{\omega_H\to 0}{\cal T}_-={\cal T}\,,
\end{equation}
as in this limit  $B({\bf x})\to I$ and $\beta ({\bf x})\to1\in\rho_D$.  Therefore ${\cal G}_+\sim{\cal G}_-$ tend to the usual discrete group of the Dirac theory in special relativity.

Trying then to continue this study with the improper isometries in physical frames we meet difficulties because we must consider the explicite transformation $\omega_H\to -\omega_H$ if we use the same coordinates on $M_+$ and $M_-$.  However, once we have investigated the physical meaning of the proper isometries in physical frames we may look for a general picture coming back to the  conformal frames where we have technical opportunities. 

\section{Discrete isometries in conformal frames}

The advantage of the conformal frames is that here  the Dirac equation as well the metric do not change their forms when we change the portion by changing the sign of $t_c$. Starting with  the proper isometries (\ref{Dp}) we observe first that the parity remains the same, 
\begin{equation}\label{cP}
	{\cal P}\psi(t_c,{\bf x}_c)=\gamma^0 \psi(t_c,-{\bf x}_c)\,,
\end{equation}
as the three-vectors ${\bf x}_c$ and ${\bf x}$ are parallel to ${\bf z}$.  The isometry $\phi^c_{[0]}$ defined by Eq. (\ref{Pi0}) mixes the time and the Cartesian coordinates such that now we may write  
\begin{equation}\label{cB}
	{\cal B}\psi (t_c,{\bf x}_c)=	\lambda^c_{[0]}({x}_c)\psi(t_c'({x}_c), {\bf x}_c'(x_c))=\gamma^0\gamma^5\beta^c(x_c)\psi(t'_c(x_c), {\bf x}'_c(x_c)) \,,
\end{equation}
where the functions $t'_c(x_c)$ and ${\bf x}'_c(x_c)$ are defined in Eq. (\ref{Pi0}). The matrix   $\beta^c=\beta\circ\chi^{-1}$ and the matrix transforming tetrads,  $\Lambda^c_{[0]}=TB^c=TB\circ \chi^{-1}$, are obtained after substituting the physical coordinates according to Eq. (\ref{subs}). A few manipulation carries out the matrix elements of $B^c(x_c)$ that read 
\begin{eqnarray}
	\left<B^c({x}_c)\right>^{0\,\cdot}_{\cdot\,0}&=&\frac{t_c^2+{\bf x}_c^2}{t_c^2-{\bf x}_c^2}\,,	\nonumber\\ 
	\left<B^c({x}_c)\right>^{0\,\cdot}_{\cdot\,i}&=&\left<B^c({x}_c)\right>^{i\,\cdot}_{\cdot\,0}=-\frac{2 x_c^i}{t_c^2-{\bf x}_c^2}\,,\label{Bc}\\
	\left<B^c({x}_c)\right>^{i\,\cdot}_{\cdot\,j}&=&\delta^i_j+\frac{2 x_c^i x_c^j}{t_c^2-{\bf x}_c^2}\,,\nonumber
\end{eqnarray}
and the new form 
\begin{equation}\label{bec}
	\beta^c(t_c,{\bf x}_c)=\frac{t_c- \gamma^0\gamma^i x_c^i}{\sqrt{t_c^2-{\bf x}_c^2}}\in \rho_D\,,	
\end{equation}
of the associated matrix through the canonical homomorphism. Finally we derive the transformations produced by the isometry $\phi^c_{{\frak p}[0]}$,
\begin{equation}\label{cPB}
	{\cal PB}\psi (t_c,{\bf x}_c)=	\gamma^5\beta^c(t_c,-{\bf x}_c)\psi(t'_c(x_c), -{\bf x}'_c(x_c)) \,,
\end{equation} 
and $\Lambda^c_{{\frak p}[0]}(x_c)=-B^c(t_c,-{\bf x}_c)$. Thus all the results concerning the proper isometries are in accordance to those obtained in physical frames.

For studying the improper isometries (\ref{Dm}) we start with the antipodal isometry $\phi^c_{\frak{a}}$
for which we have $\Lambda_{\frak a}^c=P \to \lambda^c_{\frak a} =\gamma^0$ such that by denoting $\mathsf{T}_{\frak a}={\cal A}$ we obtain the antipodal transformation of the Dirac field, 
\begin{equation}\label{cA}
	{\cal A}\psi(t_c,{\bf x}_c)=\gamma^0 \psi(-t_c,{\bf x}_c)\,,
\end{equation} 
with the help of which we may deduce the actions of the other improper isometries. Indeed, observing that $\phi^c_{[0][4]}=\phi^c_{\frak p}\circ\phi^c_{\frak a}$ we obtain its action
\begin{equation}\label{cPA}
{\cal PA}\psi(t_c,{\bf x}_c)= \psi(-t_c,-{\bf x}_c)\,,
\end{equation} 
and the  associated matrix  $\Lambda_{[0][4]}=I$. Similarly we find the action of the isometry (\ref{Pi4}) which can be decomposed as $\phi^c_{[4]}=\phi_{\frak p}\circ\phi^c_{\frak a}\circ\phi^c_{[0]}$ such that we may write 
\begin{equation}\label{cPAB}
	{\cal PAB}\psi (t_c,{\bf x}_c)=	\gamma^0\gamma^5\beta^c(-t_c,-{\bf x}_c)\psi(-t'_c(x_c), -{\bf x}'_c(x_c)) \,,
\end{equation}
deducing that the tetrad fields are transformed by $\Lambda^c_{[4]}=\Lambda^c_{[0]}=TB^c$. Hereby we find that the isometry $\phi^c_{{\frak p}[4]}=\phi^c_{\frak a}\circ\phi^c_{[0]}$ gives the transformations
\begin{equation}\label{cAB}
	{\cal AB}\psi (t_c,{\bf x}_c)=	\gamma^5\beta^c(-t_c,{\bf x}_c)\psi(-t'_c(x_c), {\bf x}'_c(x_c)) \,,
\end{equation}
and   $\Lambda^c_{{\frak p}[4]}(x_c)=-B^c(-t_c,{\bf x}_c)$. 

	\begin{table}
\begin{tabular}{ccccccc}
${\frak d}$&$T_{\frak d}$&$\phi^{c\, 0}_{\frak d}(x_c)$&$\phi^{c\, i}_{\frak d}(x_c)$&$\lambda^c_{\frak d}(x_c)$&$\Lambda^c_{\frak d}(x_c)$&Eq.\\	
&&&&&&\\
${\frak p}$&${\cal P}$&$t_c$&$-{x}_{c}^{i}$&$\gamma^0$&$P$&(\ref{cP})\\
$[0]$&${\cal B}$&$t'_c(x_c)$&${x}^{\prime\, i}_c(x_c)$&$\gamma^0\gamma^5\beta^c(t_c,{\bf x}_c)$&$TB^c(t_c,{\bf x}_c)$& (\ref{cB})\\
${\frak p}[0]$&${\cal PB}$&$t'_c(x_c)$&$-{x}^{\prime\, i}_c(x_c)$&$\gamma^5\beta^c(t_c,-{\bf x}_c)$&$-B^c(t_c,-{\bf x}_c)$&(\ref{cPB}) \\
${\frak a}$&${\cal A}$&$-t_c$&$x_c^i$&$\gamma^0$&$P$&(\ref{cA}) \\
$[0][4]$&${\cal PA}$&$-t_c$&$-x_c^i$&$1\in\rho_D$&$I$&(\ref{cPA}) \\
$[4]$&${\cal PAB}$&$-t'_c(x_c)$&$-{x}^{\prime\, i}_c(x_c)$&$	\gamma^0\gamma^5\beta^c(-t_c,-{\bf x}_c)$&$TB^c(t_c,{\bf x}_c)$&(\ref{cPAB}) \\
${\frak p}[4]$&${\cal AB}$&$-t'_c(x_c)$&${x}^{\prime\, i}_c(x_c)$&$	\gamma^5\beta^c(-t_c,{\bf x}_c)$&$-B^c(-t_c,{\bf x}_c)$&(\ref{cAB}) 
\end{tabular}
\caption{Transformations induced by the discrete de Sitter  isometries in conformal frames. The functions $t'_c(x_c)$ and ${x}^{\prime\, i}_c(x_c)$ are defined in Eq. (\ref{Pi0}) while the matrices $B^c(x_c)$ and $\beta^c(x_c)$ are given by Eqs. (\ref{Bc}) and respectively (\ref{bec}). }
\end{table}	

In Tab. 3 we summarize  the  results obtained in this section listing the transformations of the Dirac field produced by the discrete isometries  in conformal frames.  The first three are due to the proper ones among them only the parity remains global as the last two ale local, depending on the matrix (\ref{bec}). The improper isometries are two global and two local depending on the same boost. All these transformations  can be obtained by using the basis formed by ${\cal P}$, ${\cal B}$ and  ${\cal A}$.  Moreover, these transformations can be composed at any time with the charge conjugation ${\cal C}$ giving rise to a large collection of discrete transformations. For example, the transformations  (\ref{Wtr}) and can  (\ref{PWtr}) be rewritten  in conformal frames as
\begin{eqnarray}
	{\cal T}_c\psi(t_c,{\bf x}_c)&=&-\beta^c(t_c,-{\bf x}_c)\gamma^1\gamma^3\psi^*(t_c'({\bf x}),{\bf x}'_c({x_c}))\,,\label{Tc}	\\
		{\cal P}{\cal T}_c\psi(t_c,{\bf x}_c)&=&-\beta^c(t_c,{\bf x}_c)\gamma^0\gamma^1\gamma^3\psi^*(t_c'({\bf x}),-{\bf x}'_c({x_c}))\,,\label{PTc}
\end{eqnarray} 
having the same form on $M_+$ and $M_-$. With their help we can construct the discrete group ${\cal G}_c$
in conformal frames which has a similar multiplication table as in Tab. 2 but with ${\cal T}_c$ instead of ${\cal T}_+$. We observe that Eqs. (\ref{Tc}) and   (\ref{PTc})  are far from an intuitive interpretation confirming again that the conformal frames are useful for calculating the action of the discrete isometries whose physical meaning can be better understood  in physical frames. On the other hand, we must stress that the transformations listed in Tab. 3 cannot be gathered in the same discrete group as the improper ones do not have a physical meaning if we accept that one cannot perform measurements simultaneously in the expanding and collapsing universes.

\section{Concluding remarks}

We presented for the first time the complete theory of the discrete transformations of the covariant Dirac field on the de Sitter manifold whose expanding and collapsing portions are models of universes. We focused on the group of discrete symmetries in the physical frames of the expanding universe which is of actual interest in cosmology pointing out that similar results on the collapsing portion can be obtained simply by substituting $\omega_H\to -\omega_H$.  Moreover, we derived all the transformations generated by the discrete isometries, including the improper ones, using conformal frames where we have obvious technical advantages but obscure physical meaning as the improper isometries do not have a physical meaning being of a mere mathematical interest. Thus we may say that this paper complete the previous studies concerning the de Sitter continuous isometries \cite{CD1,Ccov,Cquant} generating the principal invariants of the Dirac theory on the de Sitter expanding universe \cite{ES,EPL,GRG}.

However, our results differ from those obtained in the invariant approach of the de Sitter ambient where only linear representations of the group ${\rm Pin}O(1,4)$ are considered obtaining a discrete group formed exclusively by global transformations \cite{T,V}. In contrast, in our approach only the parity, antipodal isometry,   charge conjugation and their compositions remain global while all the symmetries containing  the Wigner time reversal, ${\cal T}_+$,  lead to local transformations depending on the matrix (\ref{betax}). This indicates  that on the de Sitter manifold a global time reversal cannot be defined at least in our covariant approach. 

Finally we note that the groups of the discrete symmetries on the de Sitter portions are of the same order as in the case of the Minkowski geometry because both these manifolds are of maximal symmetry. Therefore, we expect to obtain similar results on the Anti-de Sitter space-time which is the third $(1+3)$-dimensional manifold of maximal symmetry  \cite{SW}.

\end{document}